\documentclass[pra, preprint, 12pt]{revtex4-2}

\usepackage{dcolumn, bm, indentfirst}
\usepackage{amsthm,amsmath,amssymb}
\usepackage{graphicx, tikz, orcidlink}
\usepackage{hyperref}
\hypersetup{colorlinks = true, urlcolor = blue, linkcolor = blue, citecolor = blue}

\begin{document}
\title{Reinforcement Learning Assisted Quantum Simulation of Many-Body Excited States and Real-Time Dynamics}

\date{Last updated: \today}

\author{Jiaji Zhang \orcidlink{0000-0003-2978-274X}}
\email[Authors to whom correspondence should be addressed: ]{zhang.jiaji.i92@kyoto-u.jp}
\affiliation{School of Optical Information and Energy Engineering,
Wuhan Institute of Technology, Wuhan, China}

\author{Lipeng Chen \orcidlink{0009-0002-1541-8912}}
\affiliation{Zhejiang Laboratory, Hangzhou 311100, China}

\author{Carlos L. Benavides-Riveros \orcidlink{0000-0001-6924-727X}}
\email[Authors to whom correspondence should be addressed: ]{carlos.benavides@iqm.tech}
\affiliation{IQM Quantum Computers, Georg-Brauchle-Ring 23-25, 80992 Munich, Germany}

\begin{abstract}
The computation of electronic excited states and real-time quantum dynamics of many-fermion systems is among the most promising applications of near-term quantum computing. In this work, we generalize the reinforcement learning contracted quantum eigensolver (RL-CQE), previously developed for ground-state problems, to electronic excited states and real-time quantum dynamics, in which a deep
Q-network agent adaptively selects the two-body operators at each iteration, yielding more compact ans\"{a}tze and improved robustness with respect to
critical hyperparameters. A key feature of the algorithm is a scalable state representation based on
the ACSE residuals, whose dimension grows with the one-particle basis but remains independent of the number of targeted excited states. We also verify the equivalence of sign-free qubit operators in the excited-state setting, extending a result previously established for ground-state problems. Our RL-CQE for time evolution derives from a constant-scaling ansatz that represents the wave function with a fixed number of unitary transformations independent of simulation time $t$, enabled by the shared unitary structure of the purified ensemble treatment of excited states. Benchmarks on chemical systems demonstrate chemical accuracy with minimal operator counts across a range of bond lengths.
\end{abstract}

\maketitle

\section{Introduction}

Electronic excited states play a central role in a wide range of physical and chemical processes that lie beyond the reach of ground-state or equilibrium descriptions, including photoexcited dynamics in light-harvesting complexes \cite{blankenship2002, vivancos2020jacs, scholes2011natchem}, non-equilibrium charge transport driven by laser fields \cite{greenwald2020natnano, bhan2022nanolett, lawrence2026jcp, zeng2023npjcm}, and nonlinear response phenomena in advanced quantum materials \cite{yang2024science, sazhin2024natcomm, dorfman2016rmp}. Computing these states accurately poses fundamental challenges: they often exhibit strong multi-configurational character \cite{sugisaki2016jpca, shi2018prb} and near-degeneracy at conical intersections \cite{domcke2012arpc, domcke2004book}, features that render standard single-reference methods unreliable. Implementing robust excited-state solvers on near-term quantum devices therefore represents one of the most compelling and practically urgent applications of quantum algorithms.

Among the algorithms developed to date, the contracted quantum eigensolver (CQE), which is based on the contracted Schr\"{o}dinger equation, has attracted considerable attention \cite{smart2021prl, smart2022pra, wang2023pra, smart2024jctc, benavidesriveros2024njp}. A key advantage of CQE is that the wavefunction is parameterized entirely through two-body operators, yielding favorable scaling with the number of orbitals and electrons \cite{smart2021prl}. Although originally formulated for ground-state problems, CQE has been extended to excited states through a purified ensemble approach, which recasts the excited-state problem as an effective ground-state problem on an enlarged Hilbert space \cite{benavidesriveros2024njp, benavidesriveros2022prl}. While this ensemble formulation naturally inherits improvements developed for ground-state CQE, it exhibits slower convergence due to the simultaneous optimization of multiple states, and its numerical performance is sensitive to several critical hyperparameters (most notably the ensemble weight vector) which limits its applicability to generic systems \cite{benavidesriveros2024njp}.

The rapid development of artificial intelligence and deep learning has dramatically accelerated progress across the fundamental sciences \cite{mater2019jcim, karniadakis2021nrp, zhao2024advmat, dral2024chemcomm, zhang2024jpcl, alexeev2025natcomm}. Among its subfields, reinforcement learning (RL) trains an agent to make sequential decisions in complex, potentially black-box environments \cite{sutton1998book, powell2022book, wang2024tnnls, ghasemi2025arxiv}. Unlike supervised and unsupervised learning, in which a model is passively trained on fixed datasets to minimize a predefined loss, an RL agent learns actively through exploration and exploitation, maximizing a cumulative reward without requiring prior knowledge of the problem structure.
This model-free character has made RL a powerful tool across a broad range of real-world tasks, including game playing \cite{silver2016nature, silver2018science, vinyals2019nature}, robotic control \cite{kober2013ijrr, tang2025ar, li2025arxiv}, autonomous driving \cite{kiran2022tits, dinneweth2022ais, tobisawa2025alr}, and large language models \cite{ziegler2020arxiv, liu2026esa, lambert2026arxiv}.

The model-free nature of RL makes it particularly well suited for hybrid quantum-classical algorithms, where the quantum device can naturally be treated as a black-box environment whose internal details need not be explicitly modeled \cite{alexeev2025natcomm}. RL has already been applied in this context to quantum eigensolvers and real-time quantum control \cite{yao2021prx, pai2022pra, wang2025pra, bukov2018prx, fosel2018prx, niu2019npjqi, reuer2023natcomm}, and has been shown to discover optimal operator sequences requiring fewer circuit elements than conventional optimization methods \cite{yao2021prx}. Most recently, Wang and Mazziotti \cite{wang2025pra} proposed an RL-CQE algorithm for ground-state problems, reformulating the CQE update procedure as a Markov decision process and training a deep Q-network to select two-body operators at each iteration, demonstrating significant reductions in circuit depth relative to conventional CQE implementations. Despite these advances, the application of RL to the direct optimization of excited-state wavefunctions within a quantum eigensolver framework has, to the best of our knowledge, not yet been explored.

In this work, we generalize the RL-CQE algorithm to excited-state problems of quantum chemistry Hamiltonians, building on the purified ensemble approach in which the wavefunction is parameterized as a sequence of unitary transformations applied to a set of orthonormal reference states \cite{benavidesriveros2022prl}. While sharing the same conceptual foundation as the ground-state RL-CQE of Ref.~\cite{wang2025pra}, our formulation introduces several key technical advances tailored to the excited-state setting.
Specifically, it substantially reduces per-iteration measurement overhead, provides a scalable state representation whose complexity is independent of the number of targeted excited states, and yields solutions that are markedly more robust with respect to the ensemble weight vector. In addition, we establish for the first time the equivalence of sign-free qubit operators (previously validated only for ground-state problems \cite{smart2022pra}) in the excited-state setting. We benchmark the algorithm on $\mathrm{H_2}$ and linear $\mathrm{H_3^+}$, demonstrating compact solutions with minimal operator counts while maintaining chemical accuracy across a range of bond lengths.

We further extend the RL-CQE framework to the quantum simulation of time-dependent wavefunctions, a long-standing challenge under limited quantum resources. Conventional approaches decompose the time-evolution operator into sequential short-time propagators, leading to circuit depths that grow without bound as the simulation time increases. To overcome this limitation, we introduce an algorithm that maintains a fixed ansatz size regardless of simulation time. The key insight is to expand the time-dependent wavefunction in the basis of Hamiltonian eigenstates, which are naturally shared across all targeted states within the purified ensemble framework. This shared structure allows the time-dependent wavefunction to be expressed as a fixed set of unitary transformations acting on a time-dependent reference state, which is itself compactly represented with the aid of RL. The resulting ansatz achieves a constant total operator count as a function of simulation time, opening new possibilities for scalable many-body dynamics on near-term quantum hardware.

The remainder of this paper is organized as follows. Section~\ref{sec.theory} introduces the theoretical background, covering the CQE formalism, its extension to excited states via the purified ensemble approach, the RL agent architecture, and the generalization to time evolution. Section~\ref{sec.result} presents numerical results on the benchmark systems.
Section~\ref{sec.conclusion} summarizes the main findings and outlines directions for future research.

\section{Theory}
\label{sec.theory}

We introduce the generalization of RL-CQE to excited state problem in Section {\ref{sec.theory.cqe}},  present the deep-Q network in {\ref{sec.theory.dqn}}, 
which are utilized as the main body of agent in this work, 
and discuss the generalization to the time evolution problem in Section {\ref{sec.theory.evolve}}.

\subsection{RL-CQE for Excited States}
\label{sec.theory.cqe}

We consider an electronic system described by the Hamiltonian
\begin{equation}
\hat{H} = \sum_{pq} h_{pq} \hat{c}_{p}^{\dagger} \hat{c}_{q} +
\frac{1}{2} \sum_{pqkl} g_{pq,lk} \hat{c}_{p}^{\dagger}
\hat{c}_{q}^{\dagger} \hat{c}_{k} \hat{c}_{l},
\end{equation}
where $\hat{c}_{p}^{\dagger}$ and $\hat{c}_{p}$ are the fermionic creation
and annihilation operators for the $p$-th spin-orbital, and $h_{pq}$ and
$g_{pq,kl}$ are the one- and two-electron integrals, respectively.
The $\nu$-th eigenstate $|\psi_{\nu}\rangle$ can be obtained by enforcing
the anti-Hermitian contracted Schr\"{o}dinger equation
(ACSE)~\cite{smart2021prl},
\begin{equation}
\langle \psi_{\nu} | \big[ \hat{\Gamma}_{kl}^{pq},\, \hat{H} \big] | \psi_{\nu} \rangle = 0,
\end{equation}
where $\hat{\Gamma}_{kl}^{pq} = \hat{c}_{p}^{\dagger} \hat{c}_{q}^{\dagger}
\hat{c}_{k} \hat{c}_{l}$ is the two-body operator. To solve the ACSE iteratively, we adopt an ansatz of the form~\cite{smart2021prl, wang2025pra}
\begin{equation}
|\psi_{\nu}^{(n+1)}\rangle = e^{\,\theta^{(n)} \hat{A}^{(n)}} |\psi_{\nu}^{(n)}\rangle,
\label{eq.ansatz_exp_def}
\end{equation}
where $\hat{A}^{(n)}$ is an anti-Hermitian operator and $\theta^{(n)}$ is a
real-valued scalar at the $n$-th iteration. The initial reference states at $n = 0$ are chosen to be mutually orthogonal Hartree--Fock states satisfying $\langle\psi_{\mu}^{(0)}|\psi_{\nu}^{(0)}\rangle = \delta_{\mu\nu}$, a property that is preserved by all subsequent unitary transformations.

To target multiple eigenstates simultaneously, the Rayleigh--Ritz variational
principle provides a variational lower bound for the weighted ensemble
energy~\cite{benavidesriveros2024njp},
\begin{equation}
\sum_{\nu=0}^{K-1} w_{\nu}
\langle \psi_{\nu} | \big[ \hat{\Gamma}_{kl}^{pq},\, \hat{H} \big] | \psi_{\nu} \rangle \geq 0,
\label{eq.variation}
\end{equation}
where the weights $\{w_{\nu}\}$ are a fixed set of positive values that are
normalized, $\sum_{\nu=0}^{K-1} w_{\nu} = 1$, and strictly decreasingly
ordered, $w_{\nu} \geq w_{\nu+1} > 0$.

The excited-state CQE with exponential ansatz maps naturally onto an Markov decision process, which lies in the theoretical foundation of RL. Within this framework, the CQE update procedure and its associated quantum algebra are regarded as the RL environment.  The anti-Hermitian operators $\hat{A}^{(n)}$ applied at each updating iteration are regarded as RL action. In this work, we utilize the sign-free two-qubit operators as the RL action, defined bin Ref.~{\cite{smart2022pra}}:  
\begin{equation}
\hat{\gamma}_{kl}^{pq} = \hat{\sigma}_{p}^{+} \hat{\sigma}_{q}^{+} \hat{\sigma}_{k}^{-} \hat{\sigma}_{l}^{-},
\end{equation}
where $\hat{\sigma}_{p}^{\pm} = \hat{\sigma}_{p}^{x} \pm i \hat{\sigma}_{p}^{y}$ is the raising/lowering operator on the $p$-th qubit, with $\hat{\sigma}^{x}$ and $\hat{\sigma}^{y}$ the standard Pauli matrices. These operators differ with the original two-body operators $\hat{\Gamma}_{k l}^{p q}$ up to a non-local sign factor that arises from the encoding of fermionic operators as qubits {\cite{nielsen2012book, bravyi2002aop}} .  
Although $\hat{\gamma}_{kl}^{pq}$ does not strictly respect fermionic statistics, it has been demonstrated to yield the same convergence behavior as  the original fermionic operators in the ground state problem {\cite{smart2022pra}}.  Here we extend this equivalence to the more generalized excited state cases. 

At each updating iteration, an RL agent follows a policy that will be parameterized as a neural network to select one of the operators $\hat{\gamma}_{k l}^{pq}$ as the RL action based on the RL state. 
The RL state, which fully characterizes the CQE, is represented by the vector of CSE residuals $\bm{r}^{(n)} = \{r_{pq,kl}^{(n)}\}$, which is defined through $\hat{\gamma}_{kl}^{pq}$ as
\begin{equation}
r_{p q, kl}^{(n)} = \sum_{\nu = 0}^{K-1} w_{\nu}\langle \psi_{\nu}^{(n)} 
|\big[ \hat{\gamma}_{k l}^{p q}, \hat{H} \big]  | \psi_{\nu}^{(n)} \rangle .
\label{eq.acse_residual}
\end{equation}
Crucially, the dimension of $\bm{r}^{(n)}$ depends only on the one-particle basis size and is entirely independent of the number of excited states $K$. This is a non-trivial scalability advantage over any representation based directly on the $K$ wavefunctions $\{|{\phi}_{\nu}^{(n)}\rangle\}$, whose combined description would grow with $K$. In addition, the variational principle for excited states, defined in  Eq.{\eqref{eq.variation}}, can be generalized to the sign-free operators $\hat{\gamma}_{k l}^{p q}$, as already shown in the ground state problem {\cite{smart2022pra}}, which provides a variational lower bound for CSE residuals. Once the action $\hat{\gamma}^{(n)}$ has been selected, we solve the scalar factor $\theta^{(n)}$ from the one-dimensional optimization,
\begin{equation}
\theta^{(n)} = \operatorname{argmin}_{\theta} \sum_{\nu=0}^{K-1} w_{\nu} 
\langle {\psi}_{\nu}^{(n)}|e^{-\theta \hat{A}^{(n)}} \hat{H} 
e^{\theta \hat{A}^{(n)}} | {\psi}_{\nu}^{(n)} \rangle,
\label{eq.theta_optim}
\end{equation}
with $\hat{A}^{(n)} = \hat{\gamma}^{(n)} - \hat{\gamma}^{(n) \dagger}$ to ensure unitarity.

\subsection{Deep-Q Network}
\label{sec.theory.dqn}

The RL policy introduced in the previous subsection is parameterized using a deep neural network, placing our algorithm within the framework of deep reinforcement learning~\cite{wang2024tnnls}. Specifically, we adopt the deep Q-network (DQN)~\cite{mnih2013arxiv, mnih2015nature, wang2016jmlr}, one of the most successful value-based deep RL algorithms, which has already been applied to quantum computing tasks~\cite{yao2021prx, pai2022pra, wang2025pra}. DQN employs a neural network to approximate the state-action value function, also known as the Q-function, within the standard Q-learning framework.

We denote by $\pi(a_n \mid s_n)$ the probability of selecting action $a_n$ when the environment is in state $s_n$ at step $n$. The corresponding Q-function is defined as the expected cumulative reward under policy $\pi$,
\begin{equation}
Q(s, a) = \mathbb{E}_{\pi} \!\left[
R_{n+1} + \eta\, Q(s_{n+1}, a_{n+1})
\mid s_{n} = s,\; a_{n} = a
\right],
\label{eq.bellman_action}
\end{equation}
where $R_n$ is the immediate reward at step $n$ and $0 < \eta < 1$ is a discount factor that controls the weight assigned to future rewards and ensures convergence of the recursion. The Q-function is updated iteratively during training according to
\begin{equation}
Q(s_t, a_t) \;\leftarrow\; Q(s_t, a_t) +
\alpha \!\left[
R_{t+1} + \gamma \max_{a}\, Q(s_{t+1}, a) - Q(s_t, a_t)
\right],
\label{eq.bellman_update}
\end{equation}
where $\alpha$ is the learning rate. In DQN, the Q-function is approximated by a neural network $Q_{\bm{\theta}}(s, a)$ with learnable parameters $\bm{\theta}$, trained by minimizing the mean-squared Bellman error implied by Eq.~\eqref{eq.bellman_update} via gradient descent.

The reward signal is defined as
\begin{equation}
R_{n} = -\sum_{\nu=0}^{K-1} w_{\nu}
\langle \psi_{\nu}^{(n)} | \hat{H} | \psi_{\nu}^{(n)} \rangle
- \lambda\, \|\bm{r}^{(n)}\|,
\label{eq.reward}
\end{equation}
where $\|\cdot\|$ denotes the Euclidean norm of the residual vector, and $\lambda \geq 0$ is an adjustable regularization parameter that controls the relative weight between energy minimization and residual suppression.

\subsection{RL-CQE for Time Evolution}
\label{sec.theory.evolve}

Encoding the time evolution $|\Psi(t)\rangle = e^{-i\hat{H}t}|\Psi(0)\rangle$ as a sequence of unitary transformations has long been a fundamental challenge in quantum simulation. Conventional algorithms based on Trotterization decompose the propagator into sequential short-time unitaries, leading to circuit depths that can grow without bound as the simulation time increases. In this subsection, we introduce a time-evolution algorithm based on the excited-state RL-CQE framework that requires only a fixed number of unitaries, so that the total computational cost remains constant in $t$.

We begin from the iterative ansatz of Eq.~\eqref{eq.ansatz_exp_def} and assume that the Hamiltonian eigenstates can be obtained by sequentially applying $N$ unitary transformations to the reference Hartree--Fock states,
$|\psi_{\nu}\rangle = \hat{U}_{N} \cdots \hat{U}_{1} |\psi_{\nu}^{(0)}\rangle$, where $\hat{U}_{n}$ is the $n$-th unitary defined by the right-hand side of Eq.~\eqref{eq.ansatz_exp_def}. We then expand the time-dependent wavefunction in the basis of Hamiltonian
eigenstates,
\begin{equation}
|\Psi(t)\rangle
= \sum_{\nu=0}^{K-1} c_{\nu}(t)\, |\psi_{\nu}\rangle
= \hat{U}_{N} \cdots \hat{U}_{1}\, |\phi^{(0)}(t)\rangle,
\label{eq.evolve_eigen}
\end{equation}
where
\begin{equation}
c_{\nu}(t) = e^{-iE_{\nu}t}\langle\psi_{\nu}|\Psi(0)\rangle
\end{equation}
are the time-dependent expansion coefficients, and
\begin{equation}
|\phi^{(0)}(t)\rangle = \sum_{\nu=0}^{K-1} c_{\nu}(t)\,|\psi_{\nu}^{(0)}\rangle
\label{eq.phi0_def}
\end{equation}
is a time-dependent superposition of the reference Hartree--Fock states. Equation~\eqref{eq.evolve_eigen} is the central object of our time-dependent treatment: because all eigenstates can be obtained simultaneously using the \emph{same} set of unitary transformations $\{\hat{U}_n\}$, that shared set can be reused directly in the construction of the time-dependent wavefunction. This is in sharp contrast to state-specific treatments of excited states, in which each eigenstate is associated with a different set of unitaries and orthogonality is enforced by additional penalty terms. The number of required unitaries $N$ is constant in $t$ and is determined
solely by the number of eigenstates $K$ that have non-zero overlap with the initial state $|\Psi(0)\rangle$.

The ansatz in Eq.~\eqref{eq.evolve_eigen} requires prior knowledge of all relevant Hamiltonian eigenstates and the explicit construction of the
reference state $|\phi^{(0)}(t)\rangle$, which is, as a result, inconvenient in practice. To arrive at a more practical form, we introduce a secondary set of time-dependent unitaries that prepares $|\phi^{(0)}(t)\rangle$ from a simple fixed reference state, chosen here to be the Hartree--Fock ground state $|\phi_0\rangle$,
\begin{equation}
|\phi^{(0)}(t)\rangle =
\hat{U}_{N^{\prime}}^{\prime}(t) \cdots \hat{U}_{1}^{\prime}(t)\,|\phi_0\rangle.
\label{eq.secondary_ansatz}
\end{equation}
Because $|\phi^{(0)}(t)\rangle$ is a superposition of simple Hartree--Fock states, the number of secondary unitaries $N^{\prime}$ is only affected by the number $K$, but is nearly independent of $t$.
Combining Eqs.~\eqref{eq.evolve_eigen} and~\eqref{eq.secondary_ansatz}, we conclude that any $|\Psi(t)\rangle$ can be prepared from the fixed reference $|\phi_0\rangle$ by a total of $N + N^{\prime}$ unitary transformations, a number that remains constant throughout the simulation.

In practice, both sets of unitaries are determined within the RL-CQE framework with a minor modification to the reward function. Adopting the same ansatz form as Eq.~\eqref{eq.ansatz_exp_def},
\begin{equation}
|\phi^{(n;\,t)}\rangle =
e^{\,\theta^{(n;\,t)}\hat{A}^{(n;\,t)}}\,|\phi^{(n-1;\,t)}\rangle,
\qquad |\phi^{(0;\,t)}\rangle \equiv |\phi_g\rangle,
\label{eq.evolve_ansatz}
\end{equation}
we proceed as follows.
Given the wavefunction $|\Psi(t)\rangle$ at time $t$, we first construct the
target state
\begin{equation}
|\chi\rangle = e^{-i\hat{H}\delta_t}\,|\Psi(t)\rangle,
\end{equation}
where $\delta_t$ is a small time step.
The operators $\hat{A}^{(n,\,t+\delta_t)}$ and scalar parameters $\theta^{(n,\,t+\delta_t)}$ are then selected by the same RL-CQE procedure detailed in the preceding subsections, including the same choice of RL state and RL action, but with the reward replaced by the fidelity,
\begin{equation}
R_{n}^{\prime} = \bigl|\langle\chi\,|\,\phi^{(n;\,t)}\rangle\bigr|^{2}.
\label{eq.fidelity_reward}
\end{equation}
The algorithm is run until the fidelity approaches unity, at which point the wavefunction at the next time step is obtained as $|\Psi(t+\delta_t)\rangle = |\phi^{(M;\,t)}\rangle$, where $M$ is the number of iterations required. 
Based on the analysis above, we can conclude that $M \le N + N^{\prime}$, and likewise remains constant in $t$. The utilization of RL allows us to obtain more efficient form with less number of unitaries. We note that the ACSE residual of Eq.~\eqref{eq.acse_residual} remains a valid state descriptor for the time-evolution problem, as established in the context of the time-dependent contracted Schr\"{o}dinger equation~\cite{nakatsuji1999tcatcm}.

Before closing this subsection, we briefly compare the present RL-CQE time-evolution algorithm with the recently developed correlation-based efficient time-evolution (CETE) algorithm of Rose et al.~\cite{rose2025arxiv}. In Ref.~\cite{rose2025arxiv}, the time-dependent contracted Schr\"{o}dinger equation~\cite{nakatsuji1999tcatcm} is adopted as the theoretical foundation, and the time-evolution problem is recast as a sequence of time-independent
problems. By employing an iterative ansatz analogous to Eq.~\eqref{eq.evolve_ansatz} and updating it via fidelity gradients in analogy with conventional CQE,
CETE achieves an ansatz whose computational cost remains constant in $t$. The present work arrives at an equivalent constant-scaling ansatz form from a different theoretical starting point, namely the purified ensemble treatment of excited states, which naturally provides a shared set of unitary transformations across all targeted eigenstates.
Beyond the theoretical difference, the RL-assisted update strategy employed here offers several practical advantages over the gradient-based approach of CETE: it requires no gradient evaluation, is less susceptible to local minima, and adaptively selects operators based on the global structure of the optimization landscape, as demonstrated for the ground- and excited-state
problems in the preceding subsections.

\section{Results and Discussion}
\label{sec.result}

In this section, we benchmark RL-CQE on the H$_2$ and linear equidistant H$_3^+$ molecules for both excited-state energies and time evolution, which
represent two of the most widely studied benchmark systems for the quantum simulation of electronic excited states~\cite{doi:10.1021/acs.jctc.4c00915}. The one- and two-electron integrals are computed using the Slater-type orbital basis STO-6G as implemented in PySCF~\cite{sun2020jcp}. The molecular Hamiltonians are mapped to 4 and 6 qubits, respectively, via the Jordan--Wigner transformation, and all quantum simulations are performed using a noiseless statevector simulator.

The DQN is implemented as a feedforward neural network with 8 linear layers, 512 hidden channels, and Gaussian error linear unit (GELU) activation
functions. Training follows the standard DQN workflow with an experience replay buffer
of capacity $10^{6}$ transitions, a discount factor $\eta = 0.99$, and a residual regularization weight $\lambda = 0.5$ in the reward function of
Eq.~\eqref{eq.reward}. The maximum number of RL steps per episode is set to 5 unless otherwise
specified. The network is optimized using the AdamW optimizer with a learning rate of $2 \times 10^{-4}$, a batch size of 256, and 3000 training episodes. All remaining hyperparameters follow the default values of the PyTorch library~\cite{paszke2019book}. The source code for this work is publicly available on GitHub.

\begin{figure}[h]
\centering
\includegraphics[width=0.7\textwidth]{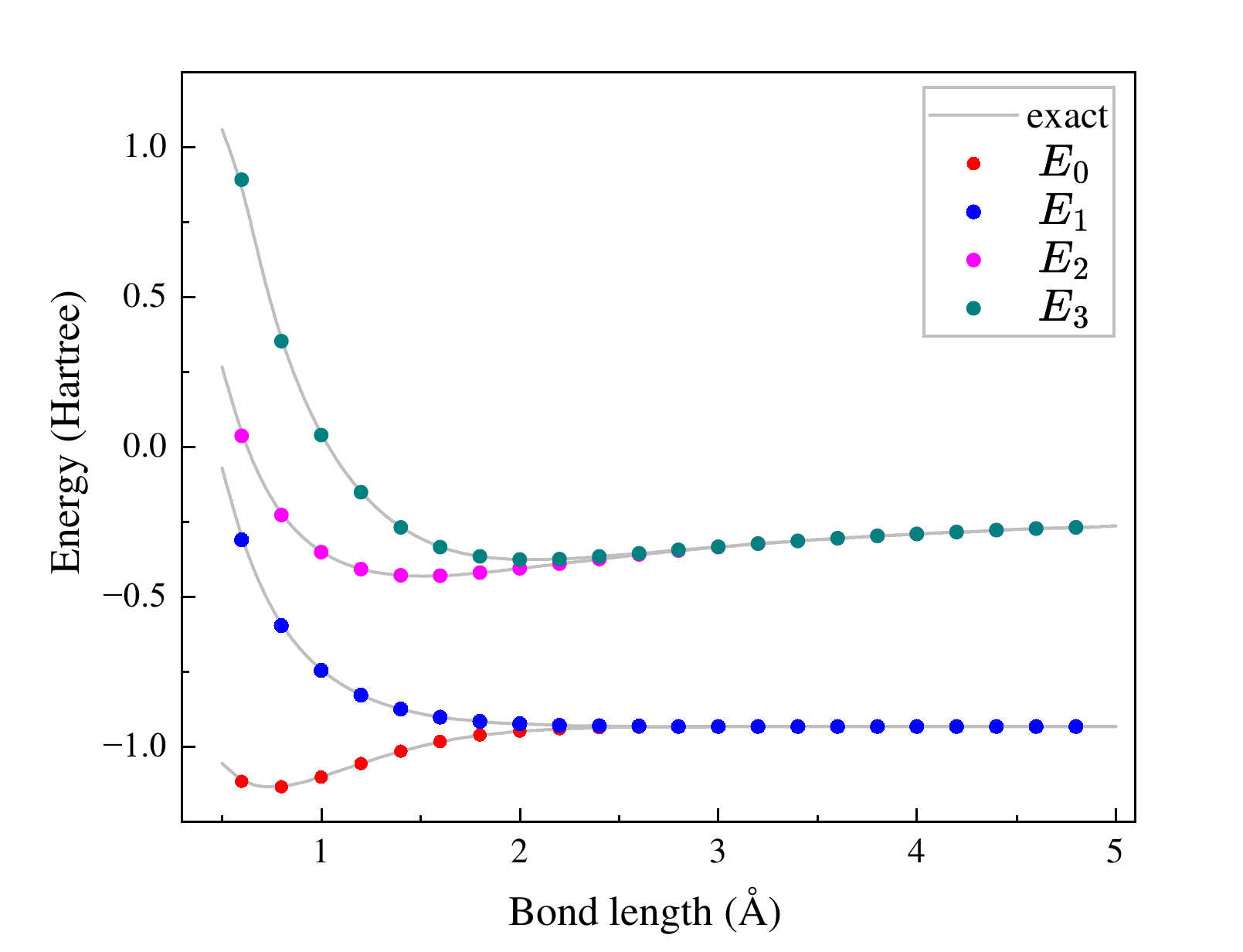}
\caption{The energy eigenvalues of ${\rm{H_2}}$ within spin sector $\langle \hat{S}^z \rangle = 0$ at bond length 0.5 ${\rm{\AA}}$ to 5.0 ${\rm{\AA}}$. The results from RL-CQE are represented by the colored dot, while the exact results from full CI are represented by the corresponding colored lines.}
\label{fig.h2_curve}
\end{figure} 

\subsection{Excited state of ${\rm{H_2}}$}

We first benchmark RL-CQE on the H$_2$ molecule, targeting the lowest four eigenstates within the spin sector $\langle\hat{S}_z\rangle = 0$. Following Ref.~\cite{benavidesriveros2024njp}, we adopt the weight vector $\bm{w} \propto [9, 9, 1, 1]$ prior to normalization. While these weights were reported to be optimal for conventional CQE, we will show that RL-CQE is largely insensitive to this choice. Figure~\ref{fig.h2_curve} shows the potential energy curves of H$_2$ at various bond lengths, with exact results obtained from full configuration
interaction (FCI) calculations. With at most 5 unitary transformations, RL-CQE achieves energies within $10^{-3}$~Hartree of the FCI reference across all geometries, demonstrating both the accuracy of the algorithm and the effectiveness of sign-free qubit operators as the building blocks of the exponential ansatz.

\begin{figure}[h]
\centering
\includegraphics[width=0.7\textwidth]{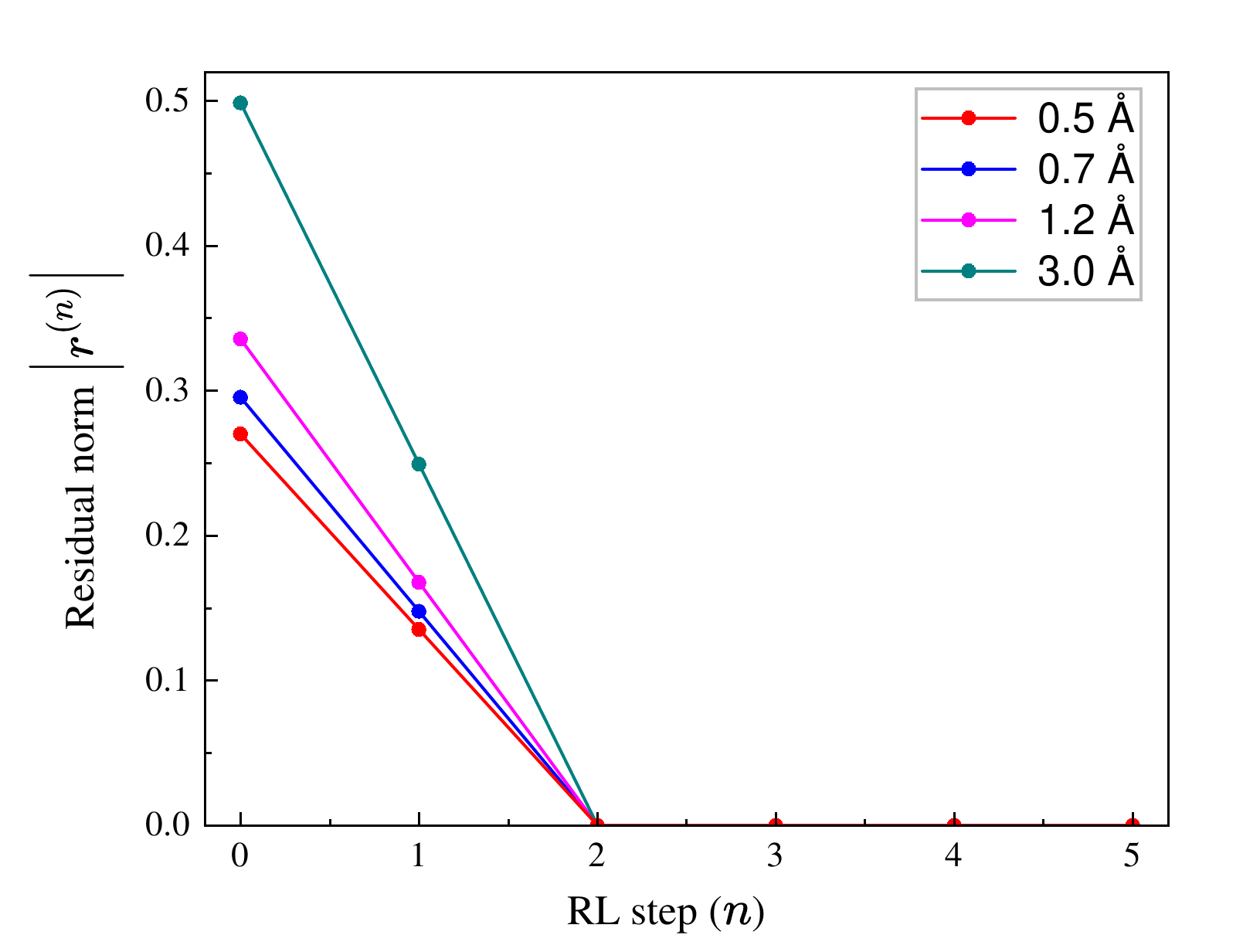}
\caption{Norm of CSE residual $|\bm{r}^{(n)}|$ against RL steps at different bond length.
All the parameters are the same as Fig.{\ref{fig.h2_curve}}. }
\label{fig.h2_steps}
\end{figure} 

One of the key features of RL-CQE is its ability to minimize the number of operators by adjusting the upper limit of RL steps. To illustrate this, we select four representative bond lengths: (i)
$0.5~\text{\AA}$ (repulsive region); (ii) $0.7~\text{\AA}$ (near equilibrium); (iii) $1.2~\text{\AA}$ (intermediate region); and (iv)
$3.0~\text{\AA}$ (dissociation limit). For all four cases, RL-CQE calculations are performed with a maximum of 2 steps, with all other settings unchanged from the previous subsection.

Figure~\ref{fig.h2_steps} shows the Euclidean norm of the ACSE residual $\|\bm{r}^{(n)}\|$ as a function of the RL step $n$, as defined in Eq.~\eqref{eq.ansatz_exp_def}. In all cases, RL-CQE reproduces the correct result using only 2 operators, and further increasing the number of steps yields no improvement, as the residual norm drops to near zero ($\sim 10^{-8}$), at which point the calculation can be terminated.

This compactness can be understood by contrasting RL-CQE with conventional CQE, which follows a fixed algorithmic update that applies all two-body operator coefficients simultaneously. This rigid update strategy may occasionally deviate from the direction of steepest descent of the energy and ACSE norm, resulting in a less compact ansatz that requires more operators. In contrast, the RL-based approach adaptively selects operators at each
step, consistently following the steepest descent direction and thereby producing near-optimal ansätze with a minimal operator count.

\begin{figure}[h]
\centering
\includegraphics[width=0.7\textwidth]{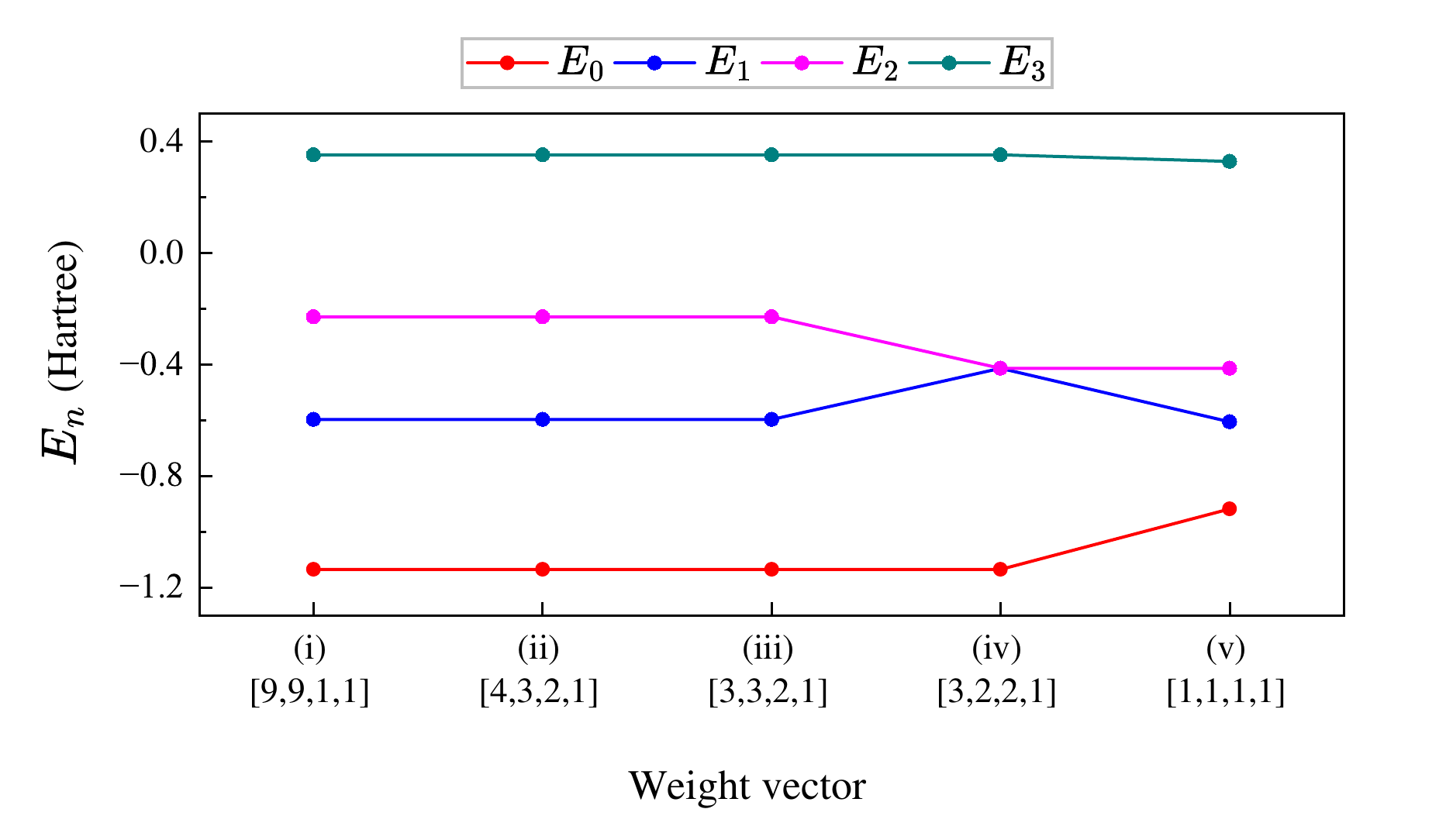}
\caption{The energy eigenvalues obtained from RL-CQE with different weight vectors.
Using weight vector (i), (ii), and (iii) lead to correct results, whereas some of the eigenvalues in (iv) and (v) are incorrect due to the improper choice of weight value.}
\label{fig.h2_weight}
\end{figure} 

Another key feature of RL-CQE is its robustness with respect to the choice of the ensemble weight vector.
In the calculations above, we adopted the weight vector $\bm{w} \propto [9, 9, 1, 1]$, which was determined by analyzing the point-group symmetry
of the system and shown to be optimal for conventional CQE. However, such optimal values are not always straightforward to determine for
more complex systems, and suboptimal choices can require significantly more operators to converge in conventional CQE.

To assess the sensitivity of RL-CQE to this choice, we fix the bond length at $0.7~\text{\AA}$ and apply RL-CQE with the following weight vectors prior to normalization: (i) $[9,9,1,1]$; (ii) $[4,3,2,1]$; (iii) $[3,3,2,1]$; (iv) $[3,2,2,1]$; and (v) $[1,1,1,1]$. The converged energy eigenvalues for each case are shown in Fig.~\ref{fig.h2_weight}.

We find that a simple strictly descending choice such as (ii) $[4,3,2,1]$ yields accuracy comparable to the optimal vector (i) $[9,9,1,1]$. Assigning identical weights to states that share a symmetry-protected degeneracy, as in case (iii) $[3,3,2,1]$, can slightly improve training stability without compromising the final accuracy. However, assigning identical weights to states that do not share such a
degeneracy, as in case (iv) $[3,2,2,1]$ where the 2nd and 3rd states receive the same weight, causes RL-CQE to converge to a mixture of the corresponding eigensubspace rather than the individual eigenstates.
In this situation, a subspace diagonalization must be performed to recover the correct eigenstates.
Similarly, the uniform vector (v) $[1,1,1,1]$ induces mixing across all eigensubspaces, leading to deviations in all eigenvalues.
These results suggest that a strictly descending weight vector such as $[4,3,2,1]$ is a robust and practical default choice for RL-CQE, requiring no prior knowledge of the system symmetry.

\subsection{Excited state of H$_3^+$}

\begin{figure}[h]
\centering
\includegraphics[width=0.7\textwidth]{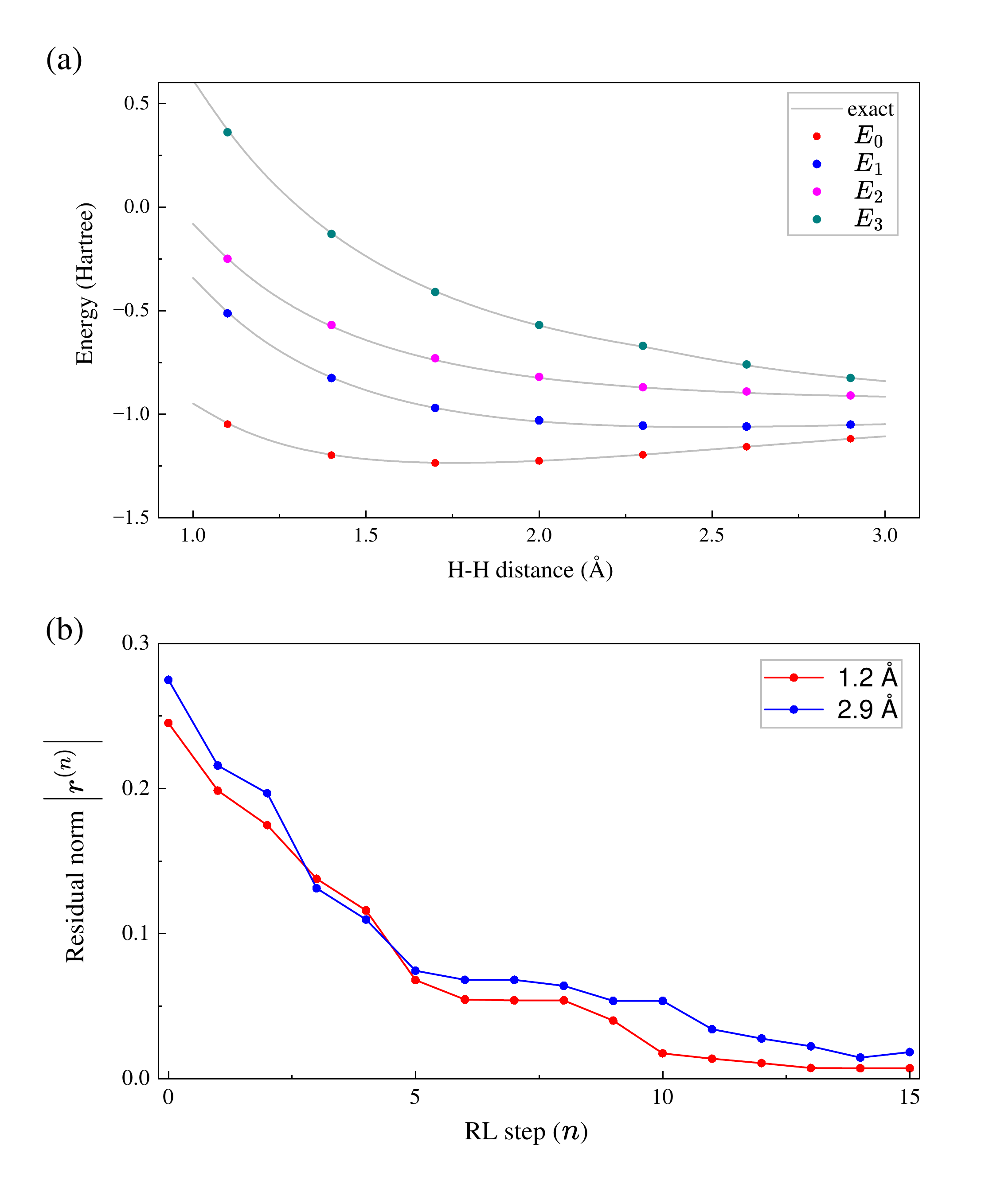}
\caption{(a) The energy eigenvalues of linear H$_3^+$ as a function of H-H distance at 1.0 ${\rm{\AA}}$ to 3.0 ${\rm{\AA}}$. 
Only lowest 4 singlet states with $\langle \hat{S}^z \rangle = 0$ are presented. 
(b) Norm of CSE residual $|\bm{r}^{(n)}|$ against RL steps at different bond length.}
\label{fig.h3_compact}
\end{figure}

We next benchmark RL-CQE on the linear equidistant H$_3^+$ molecule, a widely used test case for strong electron correlation. We target the lowest four eigenstates within the spin sector $\langle\hat{S}_z\rangle = 0$, using the weight vector $\bm{w} \propto [4,3,2,1]$ prior to normalization and a maximum of 15 RL steps. Fig.~\ref{fig.h3_compact}(a) shows the potential energy curve at various
H--H bond lengths after training for $10^{6}$ episodes, with exact results obtained from numerical diagonalization. RL-CQE yields energies in close agreement with the exact reference across all geometries. Fig.~\ref{fig.h3_compact}(b) shows the convergence of the ACSE residual norm $\|\bm{r}^{(n)}\|$ as a function of RL step $n$ at two representative geometries: (i) $1.7~\text{\AA}$ (near equilibrium) and (ii)
$2.9~\text{\AA}$ (dissociation limit).
The first five operators $\hat{\gamma}^{(1)}, \ldots, \hat{\gamma}^{(5)}$ and their corresponding parameters $\theta^{(1)}, \ldots, \theta^{(5)}$
(see Eq.~\eqref{eq.ansatz_exp_def}) are listed explicitly in Table~\ref{table.h3_seq}. In both cases, RL-CQE selects distinct operator sequences adapted to the local electronic structure at each geometry, while maintaining high accuracy and a compact operator representation in both regimes.

\begin{table}[h]
\begin{tabular}{||c | c | c | c | c | c ||} 
 \hline
RL step $(n)$ & 1 & 2 & 3 & 4 & 5   \\ 
[1ex]  \hline
1.7 ${\rm{\AA}}$, $\hat{\gamma}^{(n)}$ & $\hat{\gamma}_{56}^{12}$ &
$\hat{\gamma}_{36}^{14}$ & $\hat{\gamma}_{36}^{23}$ & 
$\hat{\gamma}_{45}^{04}$ & $\hat{\gamma}_{56}^{34} $ 
\\[1ex] \hline 
1.7 ${\rm{\AA}}$, $\theta^{(n)}$ & 0.1216 & -0.2420 & 
0.2762 & 0.1692 & 0.2631 
\\[1ex]\hline\hline
2.9 ${\rm{\AA}}$, $\hat{\gamma}^{(n)}$ & $\hat{\gamma}_{56}^{12}$ &
$\hat{\gamma}_{45}^{23}$ & $\hat{\gamma}_{45}^{14}$ & 
$\hat{\gamma}_{36}^{23}$ & $\hat{\gamma}_{56}^{34} $ 
\\[1ex] \hline 
2.9 ${\rm{\AA}}$, $\theta^{(n)}$ & 0.3758 & -0.6163 & 
0.780 & 0.3177 & 0.8851
\\[1ex]\hline
\end{tabular}
\label{table.h3_seq}
\caption{First 5 operators and factors of ansatz Eq.{\eqref{eq.ansatz_exp_def}} obtained from RL-CQE at H-H distance 1.7 ${\rm{\AA}}$ and 2.9 ${\rm{\AA}}$, respectively.}
\end{table}

\subsection{Time evolution problem}

\begin{figure}[h]
\centering
\includegraphics[width=0.7\textwidth]{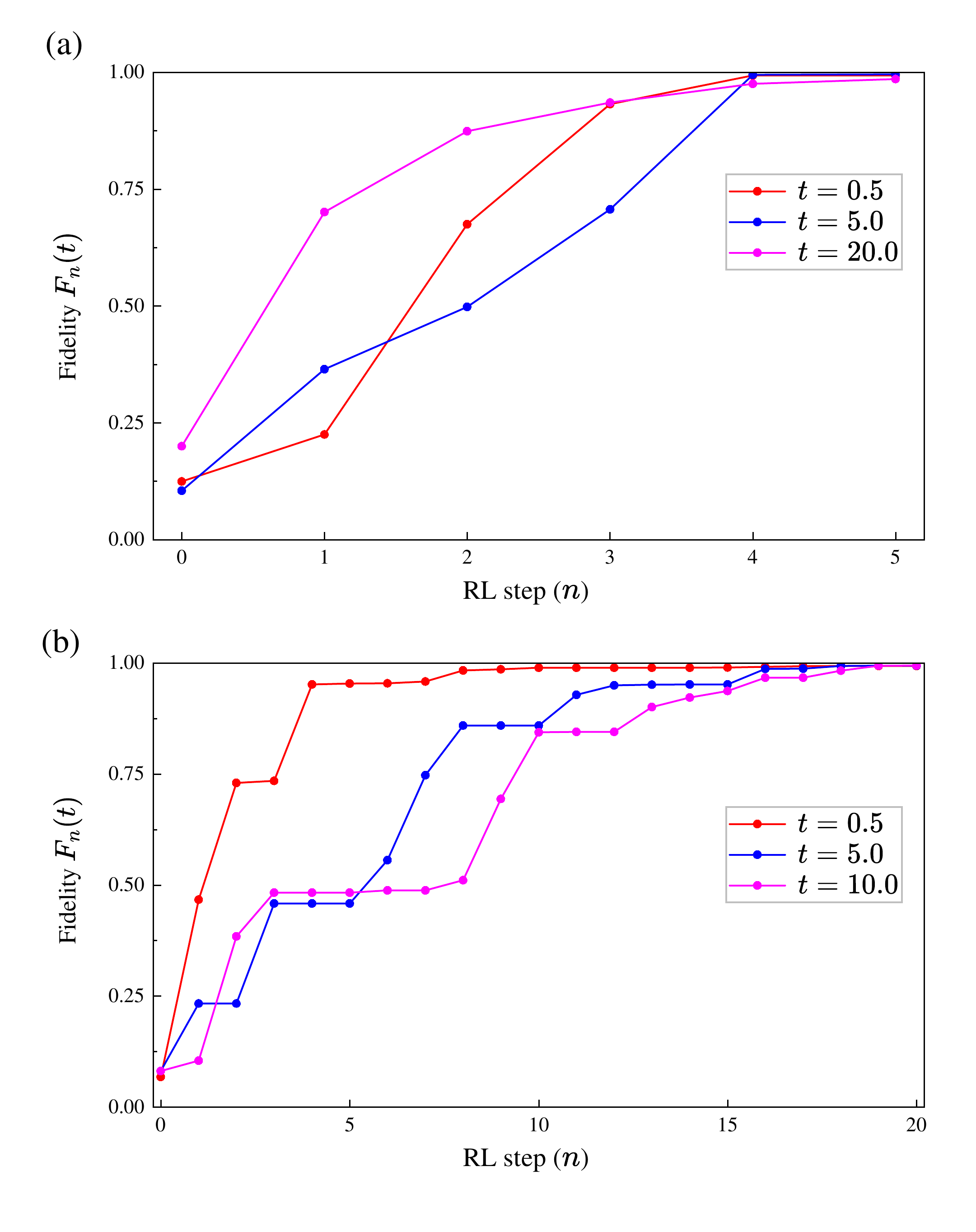}
\caption{The time-dependent fidelity $F_n(t)$ against RL step $n$ at different time points for (a) ${\rm{H_2}}$ and (b) H$_3^+$ molecules, respectively. }
\label{fig.evolve_steps}
\end{figure}

Finally, we apply RL-CQE to the time evolution of H$_2$ and H$_3^+$. The initial state $|\Psi(0)\rangle$ is chosen as a superposition of Hartree--Fock states within the spin sector $\langle\hat{S}_z\rangle = 0$ with randomly chosen coefficients. The time evolution is evaluated over $t \in [0, 20]$~a.u. with a time step $\delta_t = 0.05$~a.u., following the iterative RL-CQE procedure described in Section~\ref{sec.theory.evolve}. The reference wavefunction $|\Psi_{\rm ref}(t)\rangle$ is obtained from exact numerical diagonalization of the Hamiltonian. Fig.~\ref{fig.evolve_steps} shows the fidelity $F_n(t) = \bigl|\langle\Psi_{\rm ref}(t)|\phi^{(n;\,t)}\rangle\bigr|$, as a function of RL step $n$ at selected time points. Across the entire time range, $F_n(t)$ converges to near unity within (a) 5 steps for H$_2$ and (b) 20 steps for H$_3^+$. Crucially, the number of steps required remains constant throughout the simulation, and is comparable to that needed for the convergence of the
excited-state calculations. We note that $F_n(t)$ exhibits sharp jumps at intermediate RL steps for
H$_3^+$, which arise because only the absolute value of the overlap is considered, with the global phase factor neglected. These results demonstrate the ability of RL-CQE to generate compact operator sequences of fixed length for time-dependent wavefunctions, consistent with the theoretical analysis of Section~\ref{sec.theory.evolve}.

\section{Conclusion}
\label{sec.conclusion}

In this work, we have generalized the reinforcement learning contracted quantum eigensolver (RL-CQE), previously developed for ground-state problems, to electronic excited states and real-time quantum dynamics. The algorithm updates the wavefunction through a sequence of two-body exponential transformations, with an RL agent selecting the operator to be applied at each step from a fixed operator pool. This one-operator-per-step strategy yields more compact ans\"{a}tze and more robust convergence with respect to the ensemble weight vector compared to conventional CQE, which updates all operator coefficients simultaneously. We have also verified the equivalence of sign-free qubit operators in the excited-state setting, extending a result previously established for ground-state problems, and introduced a scalable state representation based on the ACSE residuals whose dimension is independent of the number of
targeted states.

Building on the excited-state formulation, we have further generalized RL-CQE to the simulation of time-dependent wavefunctions. The key insight is to leverage the shared unitary structure of the purified ensemble treatment, in which all targeted eigenstates are represented by the same set of unitary transformations. This allows the time-dependent wavefunction to be expressed with a fixed number of operators independent of the simulation time $t$, in contrast to conventional Trotterization-based methods whose circuit depth grows without
bound.

Benchmarks on molecular systems across a range of bond lengths demonstrate that RL-CQE achieves chemical accuracy with minimal operator counts for both excited-state energies and time-evolved wavefunctions. The number of operators required for time evolution remains constant throughout the simulation and is comparable to that needed for the
excited-state calculations, confirming the theoretical constant-scaling prediction.
Taken together, these results demonstrate that RL provides an effective and flexible framework for minimizing quantum resource requirements in
hybrid quantum-classical algorithms, with implications for near-term quantum simulation beyond what is accessible by conventional methods.

Several directions remain open for future work. One natural extension is the generalization to arbitrary quantum state preparation~\cite{feniou2024jpcl, iaconis2024npjqi}, where designing efficient and resource-minimal protocols remains a significant challenge. The adaptive operator selection strategy of RL-CQE is well suited to address this problem, as it navigates complex optimization landscapes
without requiring gradient information or prior knowledge of the problem structure. More broadly, the present framework can in principle be extended to open quantum systems \cite{doi:10.1021/acs.chemrev.4c00428}, non-equilibrium dynamics driven by external fields, fermion-boson mixtures \cite{Warren_2025}, and larger molecular systems where conventional methods become intractable, making RL-assisted quantum eigensolvers a promising avenue for future development.

\section*{Acknowledgments}

J.Z. and L.C. acknowledge the support from the National Natural Science Foundation of China (Grant No. 22473101).
C.L.B.R. gratefully acknowledges the financial support from the Royal Society of Chemistry and the European Union’s Horizon Europe Research and Innovation program under the Marie Skłodowska-Curie Grant Agreement No. 101065295-RDMFTforbosons.

\section*{Data availability}
The data and codes that support the findings of this article are openly available at the following URL/DOI: {\href{https://github.com/jiaji-zhang/RL-CSE}{https://github.com/jiaji-zhang/RL-CSE}}. 
 All codes to reproduce, examine, and improve our proposed analysis are freely available online.


\bibliography{ref_list}

\end{document}